\documentclass[12pt]{iopart}

\usepackage[utf8]{inputenc}

\usepackage{iopams}
\usepackage{setstack}

\usepackage[colorlinks,bookmarks=false,citecolor=blue,linkcolor=blue,urlcolor=blue]{hyperref}

\usepackage{xcolor}

\usepackage{graphicx}
\graphicspath{{figures/}{/}}

\usepackage[capitalise]{cleveref}
\crefname{appendix}{}{}
\crefname{section}{Sec.}{}
\usepackage[all]{hypcap}
\usepackage{placeins}    
\usepackage{flafter}     

\usepackage{cite}

\newcommand{\ud}{\ensuremath{\mathrm{d}}}
\newcommand{\ue}{\ensuremath{\mathrm{e}}}
\newcommand{\ui}{\ensuremath{\mathrm{i}}}

\newcommand{\gnat}{\ensuremath{\gamma_{\mathrm{nat}}}}
\newcommand{\munat}{\ensuremath{\mu_{\mathrm{nat}}}}
\newcommand{\bwdset}{\ensuremath{\Gamma_{\mathrm{b}}}}
\newcommand{\fwdset}{\ensuremath{\Gamma_{\mathrm{f}}}}
\newcommand{\chsaddle}{\ensuremath{\Gamma_{\mathrm{s}}}}

\newcommand{\Imag}{\ensuremath{\mathrm{Im}}}
\newcommand{\Real}{\ensuremath{\mathrm{Re}}}
\newcommand{\vecx}{\ensuremath{\boldsymbol{x}}}
\newcommand{\vecr}{\ensuremath{\boldsymbol{r}}}
\newcommand{\Hus}{\ensuremath{\mathcal{H}}}
\newcommand{\HusVec}{\ensuremath{\mathcal{H}(\vecx)}}
\newcommand{\HusAvg}{\ensuremath{\langle \Hus \rangle_\gamma}}
\newcommand{\HusAvgVec}{\ensuremath{\langle \HusVec \rangle_\gamma}}

\newcommand{\psiSqVec}{\ensuremath{|\psi(\vecr)|^2}}

\newcommand{\psiAvgVec}{\ensuremath{\langle \psiSqVec \rangle_\gamma}}
\newcommand{\tdist}{\ensuremath{t_\epsilon}}

\usepackage[normalem]{ulem}

\begin{document}

\title[Resonance states of the three-disk scattering system]{Resonance states of
the three-disk scattering system}

\author{Jan Robert Schmidt and Roland Ketzmerick}
\address{TU Dresden, Institute of Theoretical Physics and Center for Dynamics,
01062 Dresden, Germany}

\eads{\mailto{jan\_robert.schmidt@tu-dresden.de},
      \mailto{roland.ketzmerick@tu-dresden.de}}

\vspace{10pt}

\begin{indented}
    \item \today \date{\today}
\end{indented}

\begin{abstract}
For the paradigmatic three-disk scattering system, we confirm a recent
conjecture for open chaotic systems, which claims that resonance states are
composed of two factors.
In particular, we demonstrate that one factor is given by universal
exponentially distributed intensity fluctuations.
The other factor, supposed to be a classical density depending on the
lifetime of the resonance state, is found to be very well described by a
classical construction.
Furthermore, ray-segment scars, recently observed in dielectric cavities,
dominate every resonance state at small wavelengths also in the
three-disk scattering system.
We introduce a new numerical method for computing resonances, which allows for
going much further into the semiclassical limit.
As a consequence we are able to confirm the fractal Weyl law
over a correspondingly large range.
\end{abstract}


\section{Introduction}

The structure of eigenstates, along with spectral properties, is central for the
understanding of quantum systems.
In closed chaotic quantum systems, like quantum billiards or quantum maps, their
structure is well understood.
As stated by the quantum ergodicity theorem, almost all eigenstates are
uniformly distributed on the energy shell in the semiclassical
limit~\cite{Shn1974, CdV1985, Zel1987, ZelZwo1996, BaeSchSti1998, NonVor1998}.
The statistical properties of eigenstates are well described by the random wave
model~\cite{Ber1977b, McDKau1988, AurSte1991, LiRob1994, Pro1997b}.
An exception are states showing enhanced intensities on short unstable periodic
orbits, so-called periodic-orbit scars~\cite{Hel1984, Kap1999, Ver2015}.
The number of eigenstates can be obtained from the Weyl
law~\cite{Wey1911,AreNitPetSte2009}.

For open chaotic systems with partial or full escape~\cite{LaiTel2011,
AltPorTel2013} much less is known about resonance states, see e.g.\
reviews~\cite{Non2011, Nov2013}.
They have multifractal structures in phase space and for the case of full escape
concentrate on the backward-trapped set~\cite{CasMasShe1999b, KeaNovPraSie2006,
Dya2019}.
Additionally, they strongly depend on their decay rate.
In the semiclassical limit, their structure is described by
conditionally-invariant measures~\cite{NonRub2007}, but there are infinitely
many for any decay rate.
A prominent measure is the natural measure with its corresponding natural
(classical) decay rate~\cite{PiaYor1979, KanGra1985, Tel1987, LopMar1996,
DemYou2006, AltPorTel2013}.
Resonance states with this specific decay rate are well described by the natural
measure as shown in dielectric cavities~\cite{LeeRimRyuKwoChoKim2004,
ShiHar2007, WieHen2008, ShiHenWieSasHar2009, ShiHarFukHenSasNar2010, HarShi2015,
KulWie2016, BitKimZenWanCao2020, KetClaFriBae2022} and quantum
maps~\cite{NonRub2007, ClaKoeBaeKet2018, ClaAltBaeKet2019}.
The structure of resonance states was also related to short periodic
orbits~\cite{NovPedWisCarKea2009, ErmCarSar2009, PedWisCarNov2012,
CarBenBor2016, MonCarBor2023:p}.
Another object of interest are Schur vectors determined from resonance
states~\cite{SchTwo2004, KopSch2010} which have been described by classical
densities~\cite{HalMalGra2023}.
The number of complex resonance poles (above a cutoff) in the case of full
escape scales with the Hausdorff dimension of the chaotic saddle according to
the fractal Weyl law~\cite{Sjo1990, Zwo1999, Lin2002, LuSriZwo2003, SchTwo2004,
NonZwo2005, Non2006, RamPraBorFar2009, KopSch2010, EbeMaiWun2010,
PotWeiBarKuhStoZwo2012, KoeMicBaeKet2013, CarWisErmBenBor2013,
NonSjoZwo2014, Bor2014, DyaJin2017}.
In the case of partial escape the distribution of poles has been studied under
various aspects~\cite{WieMai2008, NonSch2008, BogDubSch2008,
BogDjeDubGozLebSchUlyZys2011, GutOsi2015, SchAlt2015}.
Open chaotic quantum systems are often studied using quantum maps with
absorptive or projective openings, which may have resonance spectra with
different properties than true scattering systems and improved quantum maps are
suggested~\cite{MerShu2018, YosMerShu2023}.

Recently, a factorization conjecture for fully chaotic open systems was introduced,
that applies to resonance states with arbitrary decay
rate~\cite{ClaKunBaeKet2021, KetClaFriBae2022}:
One factor is given by universal exponentially distributed intensity
fluctuations corresponding to a complex random wave model.
The other factor depends on the decay rate and is given by some classical
conditionally-invariant measure that is suitably smoothed.
There is no semiclassical theory to derive these measures, so far.
Heuristically motivated measures for all decay rates have been suggested for the
case of full escape~\cite{ClaKoeBaeKet2018} as well as for partial escape in
maps~\cite{ClaAltBaeKet2019} and dielectric cavities~\cite{KetClaFriBae2022}.
They show good, but not perfect, agreement.
For a Baker map with local randomization the exact measure has been
derived from a random vector model~\cite{ClaKet2022}.
The two factors of the factorization conjecture identify which features of a
resonance state are universal quantum (wave) phenomena and which are system
specific with a classical (ray) origin.
In particular, this explains which features neighboring resonance states have in
common and which are individual.

A new type of scarring of resonance states along segments of rays was recently
observed in dielectric cavities~\cite{KetClaFriBae2022}.
It is unrelated to periodic-orbit scars of closed systems~\cite{Hel1984,
Kap1999, Ver2015}.
Ray-segment scars were found in every resonance state at small wavelengths
and they were conceptually explained based on the factorization conjecture.

\begin{figure}
    \centering
    \includegraphics{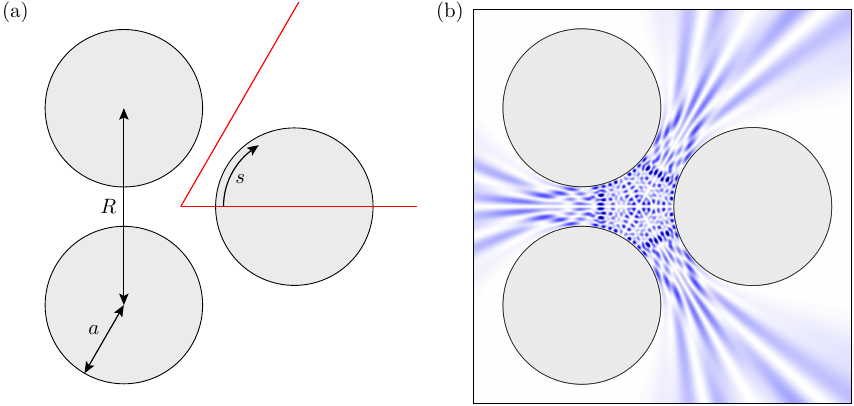}
    \caption{(a) Visualization of the three-disk scattering system with radius $a$
        of the disks and distance $R$ of their centers.
        Its fundamental domain is shown by red lines.
        The dimensionless Birkhoff coordinate $s$ along the disk's
        boundary starts at the symmetry line
        and is chosen in clockwise direction.
        (b) Exemplary resonance state in position space at
        $ka = 58.86 - 0.35 \ui$.
        Note that for clarity this figure uses $R/a = 2.5$, while
        the value $R/a = 2.1$, with disks closer to each other, is used
        throughout the paper.}%
    \label{fig:3disk_sketch}
\end{figure}

The three-disk scattering system, see \cref{fig:3disk_sketch}(a), is a paradigmatic
example since the early days of classical and quantum chaotic
scattering~\cite{Eck1987, CviEck1989, GasRic1989a, GasRic1989b, GasRic1989c,
Smi1989, EckRusCviRosSch1995, Wir1999, CviArtMaiTan2020}.
It is an autonomous system with two degrees of freedom and full escape.
More recently, the focus was on spectral properties, like the
fractal Weyl law~\cite{Lin2002, LuSriZwo2003, PotWeiBarKuhStoZwo2012,
NonSjoZwo2014, Vac2023:p},
the spectral gap~\cite{NonZwo2009, BarWeiPotStoKuhZwo2013, Vac2022:p},
and resonance chains~\cite{WeiBarKuhPolSch2014},
including experimental investigations with microwave
billiards~\cite{LuRosPanSri1999, PanLuSri2000, LuVioPanRosSri2000,
PotWeiBarKuhStoZwo2012, BarWeiPotStoKuhZwo2013}.
Resonance states were first shown in position space and the boundary phase space
in Ref.~\cite{WeiBarKuhPolSch2014}, see also \cref{fig:3disk_sketch}(b).

A semiclassical theory for resonance poles of the three-disk scattering system
is based on dynamical zeta functions consisting of periodic orbits and evaluated
using the cycle expansion~\cite{CviEck1989, EckRusCviRosSch1995,
CviArtMaiTan2020, BarSchWei2022}.
Recently, with these methods it was shown in Refs.~\cite{BarSchWei2022,
SchWeiBar2023} how one can determine semiclassical resonance states in a Husimi
representation that combines left and right resonance
states~\cite{ErmCarSar2009}.
Hence, in principle there exists a semiclassical description for resonance states of
the three-disk scattering system.
This approach has, however, the following limitations:
(i) The convergence of the cycle expansion is in practice too slow for small
wavelengths or small distances of the disks.
(ii) To date there is no semiclassical description of the (right)
resonance states fulfilling the Schrödinger equation (in contrast to the
left-right Husimi representation mentioned above).
(iii) Although individual resonance states can be computed from hundred
thousands of
periodic orbits, no insight about their structure is obtained, e.g.\ about their
dependence on the decay rate or about similarities and differences of neighboring
resonance states.
This, however, is the virtue of the factorization conjecture and it is desirable
to test it for the three-disk scattering system.

In this paper, we confirm the factorization conjecture for resonance states in
the three-disk scattering system.
In particular, we demonstrate that one factor is given by universal
exponentially distributed intensity fluctuations.
For the other factor, we show that the classical density is very well, but not
perfectly, described by extending a construction from maps with full escape to
the three-disk scattering system.
Furthermore, we observe ray-segment scars in every resonance state at small
wavelengths.
All these results on resonance states are made possible due to a new numerical
method, which allows for going about two orders of magnitude further into the
semiclassical limit than before.
This allows for confirming the fractal Weyl law over a correspondingly large range.

\section{Three-disk scattering system}

\subsection{Classical dynamics}%
\label{sec:classical_dynamics}

The three-disk scattering system
consists of three hard disks of radius $a$ with centers at the corners of an
equilateral triangle of side length $R$, see \cref{fig:3disk_sketch}(a).
It is uniquely characterized by the dimensionless ratio
$R/a$.
In this paper we focus on the parameter $R/a = 2.1$ for reasons related to the
analysis of resonance states and discussed in \cref{sec:resonance_states}.
A point particle moves along straight lines
between collisions with the disks and is specularly reflected at their boundaries.
The dynamics is chaotic with no stable periodic orbits present.
Using the systems $C_{3\mathrm{v}}$ symmetry~\cite{CviEck1993}, the dynamics can be
reduced to a fundamental domain, see \cref{fig:3disk_sketch}(a).

The phase space of the three-disk scattering system is four-dimensional
with dynamics taking place on the three-dimensional energy shell.
A further reduction is achieved by a Poincaré surface of section at the disk's
boundary, resulting in
the two-dimensional boundary phase space.
It is parametrized by dimensionless Birkhoff
coordinates $(s, p)$, where $sa$ is the arc length along the disk's boundary, see
\cref{fig:3disk_sketch}(a), and $p$ is the normalized momentum projected onto
the tangent of the boundary.
In the fundamental domain one has
$s \in [0, \pi]$ and $p \in [-1, 1]$.

\begin{figure}
    \centering
    \includegraphics{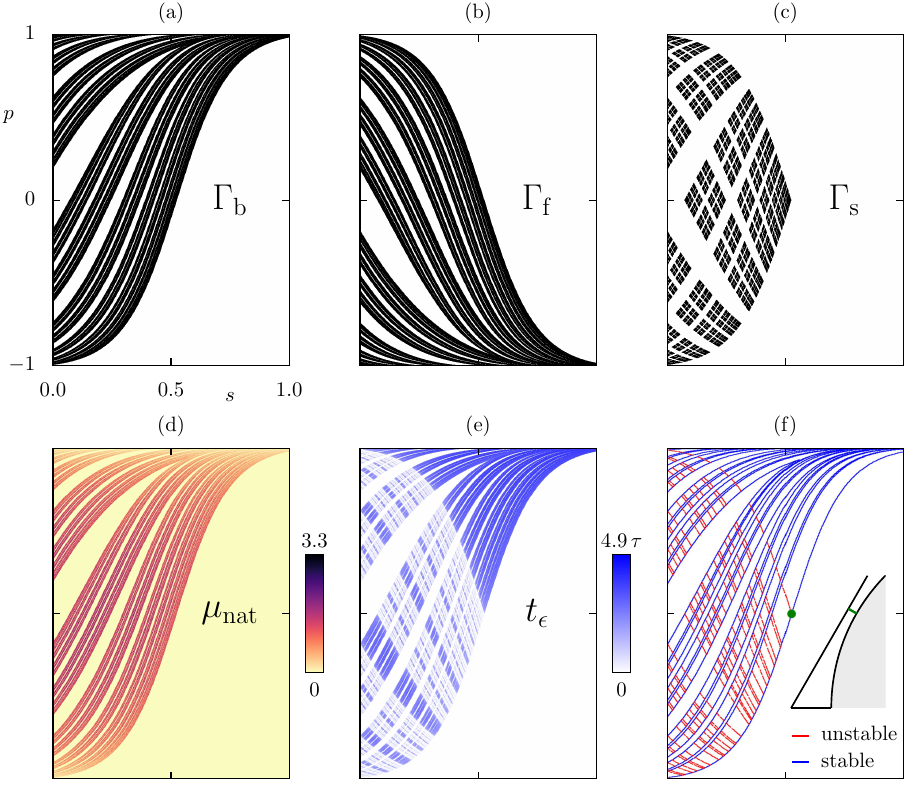}
    \caption{Invariant sets for $R/a = 2.1$ on the boundary phase space
        $(s, p)$,
        (a)~backward-trapped set $\bwdset$,
        (b)~forward-trapped set $\fwdset$, and
        (c)~chaotic saddle $\chsaddle$.
        (d)~Natural measure $\munat$ on the boundary phase space supported by
        the backward-trapped set \bwdset, where the mean value of non-zero
        pixels is set to one.
        (e)~Temporal distance $\tdist$ to an $\epsilon$-surrounding of
        the chaotic saddle $\chsaddle$ for $\epsilon = 10^{-4}$.
        (f)~Stable manifold (blue) and unstable manifold (red, restricted to $\bwdset$)
        of the periodic orbit, which has to be crossed to exit the system
        (inset).
    }%
    \label{fig:invariant_sets}
\end{figure}

Of special importance are invariant sets of the scattering system~\cite{Ott2002,
LaiTel2011, AltPorTel2013}.
The backward-trapped set $\bwdset$ consists of points that under backward time
evolution are trapped in the system.
It is shown in \cref{fig:invariant_sets}(a).
This set will be important for the localization of quantum resonance states.
The forward-trapped set $\fwdset$ is defined analogously, see \cref{fig:invariant_sets}(b).
The chaotic saddle $\chsaddle$ is given by the intersection $\bwdset \cap
\fwdset$ of backward- and forward-trapped set, see
\cref{fig:invariant_sets}(c), and thus consists of those points in phase space
which under backward and forward time evolution are trapped in the system.
These three sets are related, namely $\bwdset$ and $\fwdset$ are the unstable and
stable manifolds of $\chsaddle$,
respectively.
Numerically, they can be obtained by the sprinkler method~\cite{HsuOttGre1988,
LaiTel2011}.

These sets are known to have a multifractal structure~\cite{KanGra1985,
TelGru2006, LaiTel2011, AltPorTel2013} which is well visible in
\cref{fig:invariant_sets}.
The generalized dimension $D_q$ of the chaotic saddle in the four-dimensional
phase space can be described by $D_q = 2 d_q + 2$~\cite{AltPorTel2013}.
Here, the partial dimension $d_q$ is the generalized dimension along one
direction of the chaotic saddle in the two-dimensional boundary phase space.
We determine the box-counting dimension of the chaotic saddle in the boundary
phase space leading to the partial box-counting dimension $d_0 = 0.84$ for $R/a
= 2.1$.
This will be used for the fractal Weyl law of resonance poles, see
\cref{sec:spectrum}.

An important property of an open system is its natural measure
$\munat$~\cite{PiaYor1979, KanGra1985, Tel1987, LopMar1996, DemYou2006,
AltPorTel2013} shown in \cref{fig:invariant_sets}(d), which will be important
for the structure of resonance states, see \cref{sec:factorization_states}.
It emerges asymptotically from a uniform distribution in phase space under time
evolution leading to its concentration on the backward-trapped set $\bwdset$.
It is a conditionally-invariant measure with natural decay rate
$\gnat$.
This decay rate is also called the classical escape
rate~\cite{GasRic1989a, AltPorTel2013}, but we prefer to explicitly link
it to the natural measure, as in systems with partial escape there are
additional classically motivated rates, see e.g.\
Ref.~\cite{KetClaFriBae2022}.
The natural decay rate $\gnat$ can be expressed using the time scale $\tau = a /
v$, which is the time a particle with velocity $v$ needs to travel the distance
$a$.
For $R/a = 2.1$ we find $\gnat \tau = 0.436$.
The natural measure $\munat$ is uniformly distributed on the backward-trapped
set in phase space, while on the boundary phase space one observes a $\sqrt{1 -
p^2}$ dependence, see \cref{fig:invariant_sets}(d).
In position representation it is shown in \cref{fig:position_ray_comparison}
(top, second from left).

For the structure of resonance states it was shown in
Ref.~\cite{ClaKoeBaeKet2018} that the temporal distance $\tdist(\vecx)$ to an
$\epsilon$-surrounding of the chaotic saddle is relevant, see
\cref{sec:semiclassical_structure}.
This temporal distance $\tdist(\vecx)$, shown in
\cref{fig:invariant_sets}(e), is defined as the time a particle
starting at a phase-space point $\vecx \in \bwdset$ on the backward-trapped set
needs to come within a distance $\epsilon$ of the chaotic saddle $\chsaddle$
under backward time evolution.
In particular, on the chaotic saddle the temporal distance $\tdist$ is zero.
The temporal distance $\tdist(\vecx)$ shows a partitioning on
the backward trapped set $\bwdset$.
This partitioning can be well described
using stable and unstable manifolds of the periodic orbit
at $(s, p) = \bigl(\frac{\pi}{6}, 0\bigr)$,
which has to be crossed in position space to exit the system,
see \cref{fig:invariant_sets}(f).

\subsection{Quantization and numerical method}%
\label{sec:quantization_numerical_method}

The quantum dynamics of the three-disk scattering system is described by the
free Schrödinger equation
\begin{equation}
    \label{eq:schroedinger}
    -\Delta \psi(\vecr) = k^2 \psi(\vecr),
\end{equation}
where the wave function $\psi(\vecr)$ fulfills Dirichlet boundary conditions at
the boundary of each disk.
We focus on the antisymmetric $\mathrm{A}_2$-representation of the system's
$C_{3\mathrm{v}}$ symmetry, where Dirichlet boundary conditions are also imposed
on the symmetry lines~\cite{GasRic1989c}.
The wavenumber $k \in \mathbb{C}$ is complex with its real part being inversely
proportional to the wavelength.
Its imaginary part is proportional to the decay rate $\gamma$ of the resonance
state,
\begin{equation}
    \label{eq:decay_rate}
    \Imag \: ka = -\frac{\gamma \tau}{2},
\end{equation}
where the time scale $\tau = a / v$
is based on the $k$-dependent velocity $v = \hbar \: \Real \: k / m$ of a particle with
mass $m$.
Note that, for electromagnetic waves the constant velocity $v$ is given by the speed of
light $c$.

The physically most interesting states of the three-disk scattering system are its
resonances which consist of outgoing waves only and occur at discrete values
$k_n \in \mathbb{C}$.
The resonance poles $k_n$ are the poles of the $S$-matrix
and can be obtained from the zeros of a
matrix $M(k)$, explicitly given in \cref{sec:appendix_numerical_method},
such that $\det(M(k_n))~=~0$~\cite{GasRic1989c}.
The left singular vector of $M(k_n)$, which corresponds to the singular value zero,
allows for determining the (right) wave function $\psi_n(\vecr)$ of this
resonance state fulfilling \cref{eq:schroedinger}~\cite{WeiBarKuhPolSch2014}.
For an exemplary resonance state see \cref{fig:3disk_sketch}(b).

We are able to determine poles and resonance states up to $\Real \: ka \approx 10^5$
for $R/a = 2.1$,
which is about two orders of magnitude further in the semiclassical limit
compared to previous publications on the three-disk scattering
system~\cite{Wir1999, LuSriZwo2003, WeiBarKuhPolSch2014}.
For larger $R/a$, e.g.\ $R/a = 2.5$, we even find poles for
$\Real \: ka \approx 10^6$ (not shown).
To the best of our knowledge this goes well beyond existing numerical analysis
of any other closed or open billiard system.
This is based on three ingredients, see
\cref{sec:appendix_numerical_method} and Python code provided as
supplementary material:

\begin{enumerate}
    \item  We use that in the three-disk scattering system the application of
        the matrix $M(k)$ to a vector can be efficiently calculated using fast
        Fourier transforms, which need $\mathcal{O}(N \log{N})$ instead of
        $\mathcal{O}(N^2)$ operations.
        This allows for treating matrix dimensions $N \approx \Real \: k a$ for
        which the matrix $M(k)$ could not be stored in memory.

    \item  We find all poles near a complex wavenumber $k$ using a Taylor
        expansion of the matrix $M(k)$, extending an approach for quantum
        billiards~\cite{VebProRob2007, PeiDieHua2019} to complex $k$.
        We increase the accuracy of poles and states to a desired precision by
        subsequent convergence steps.
        Previously, this procedure was successfully applied to dielectric
        cavities~\cite{KetClaFriBae2022}.
        For these steps, the matrix-vector multiplication from (i) allows
        for employing computationally efficient iterative methods for
        eigensystems and sets of linear equations.

    \item  The calculation of wave functions $\psi(\vecr)$ and Husimi functions
        $\HusVec$ from the normal derivative of the wave function on the disk's boundary is
        substantially accelerated by using fast Fourier transforms.
\end{enumerate}

\section{Spectrum and fractal Weyl law}%
\label{sec:spectrum}

\begin{figure}
    \centering
    \includegraphics{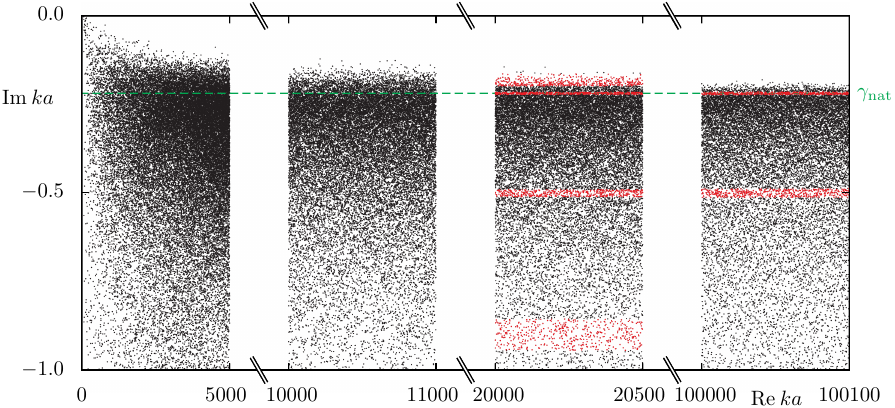}
    \caption{Spectrum for $R/a = 2.1$ for various intervals in $\Real \: ka$ and
        $\Imag \: ka \in [-1, 0]$ showing about $10^5$ resonance poles.
        The value of $\Imag \: ka = -\frac{\gnat \tau}{2}$ corresponding via \cref{eq:decay_rate} to the
        natural decay rate $\gnat$ is shown as a line (dashed).
        The poles corresponding to resonance states used for averaging in
        \cref{sec:avg_res_states} are marked in red.}%
    \label{fig:spectrum}
\end{figure}

The spectrum for $R/a = 2.1$ is obtained for $\Imag \: ka \in [-1, 0]$
and selected intervals in $\Real \: k a$
up to $ka \approx 10^5$, see
\cref{fig:spectrum}.
At the upper end of the spectrum a gap to the real line is
observed~\cite{GasRic1989b, NonZwo2009, BarWeiPotStoKuhZwo2013, Vac2022:p},
except for smaller $k$.
In particular, for $\Real \: ka < 60$ all poles are very close to the real
line as the corresponding wavelength is larger than the opening between the
disks.
For increasing $k$ the gap increases with the upper end of the spectrum
converging towards $\Imag \: ka = -\frac{\gnat \tau}{2}$, see
\cref{eq:decay_rate}.
This is in agreement with a recent conjecture about the asymptotic
spectral gap~\cite{JakNau2012, FauWei2017}.
For larger $|\Imag \: ka|$ the density decreases as expected from the analogy to
truncated random matrices~\cite{ZycSom2000, SchTwo2004, Bog2010}.

\begin{figure}
    \centering
    \includegraphics{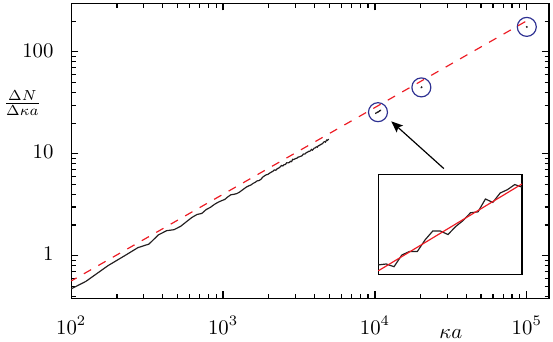}
    \caption{Fractal Weyl law for density of states,
        \cref{eq:fractal_weyl_dos}, approximated with bin size $\Delta \kappa a =
        50$ and cutoff $Ca = 1$ for the poles shown in \cref{fig:spectrum}.
        A power-law fit in the interval $\kappa a \in [10^2, 10^5]$ yields
        the power-law exponent $\delta = 0.85$.
        The fit function is shown in the main panel (shifted, dashed line) and
        in the inset for $\kappa a \in [10000, 11000]$ (not shifted, solid
        line).
        The circles highlight the density of states in the smaller intervals at
        larger values of $\kappa a$.}%
    \label{fig:fractal_weyl_law}
\end{figure}

This unprecedented large number of resonance poles allows for a comparison
with the fractal Weyl law~\cite{Sjo1990, Zwo1999, Lin2002, LuSriZwo2003,
SchTwo2004, NonZwo2005, Non2006, RamPraBorFar2009, KopSch2010, EbeMaiWun2010,
PotWeiBarKuhStoZwo2012, KoeMicBaeKet2013, CarWisErmBenBor2013, NonSjoZwo2014,
Bor2014, DyaJin2017}.
It states that the number $N(\kappa)$ of resonances $k_n$ with real part up to
$\kappa \in \mathbb{R}$ and with an imaginary part above some threshold $-C$
scales as
\begin{equation}
    \label{eq:fractal_weyl_law}
    N(\kappa) =
    \#\{k_n \: | \: \Imag \: k_n > -C \: \land \: \Real \: k_n \leq \kappa \}
    \sim \kappa^{d_\mathrm{H} + 1},
\end{equation}
where $d_\mathrm{H}$ is the partial Hausdorff dimension of the chaotic saddle.
As we computed the spectrum in intervals only, see \cref{fig:spectrum},
we cannot determine $N(\kappa)$.
Instead, we compare to the fractal Weyl law for the density of states,
\begin{equation}
    \label{eq:fractal_weyl_dos}
    \frac{\ud N(\kappa)}{\ud \kappa} \sim \kappa^{d_\mathrm{H}}.
\end{equation}
This is approximated by $\frac{\Delta N}{\Delta \kappa a}$ using $\Delta \kappa
a = 50$ in \cref{fig:fractal_weyl_law} where we find a power law $\kappa^\delta$ with
$\delta = 0.85$ up to $\kappa a = 10^5$.
This is in agreement to the partial box-counting dimension $d_0 = 0.84$ of the
chaotic saddle, see \cref{sec:classical_dynamics}, which is used to
approximate the partial Hausdorff dimension $d_\mathrm{H}$ appearing in the
fractal Weyl law.

\section{Factorization of resonance states}%
\label{sec:factorization_states}

\subsection{Resonance states}%
\label{sec:resonance_states}

\begin{figure}
    \centering
    \includegraphics{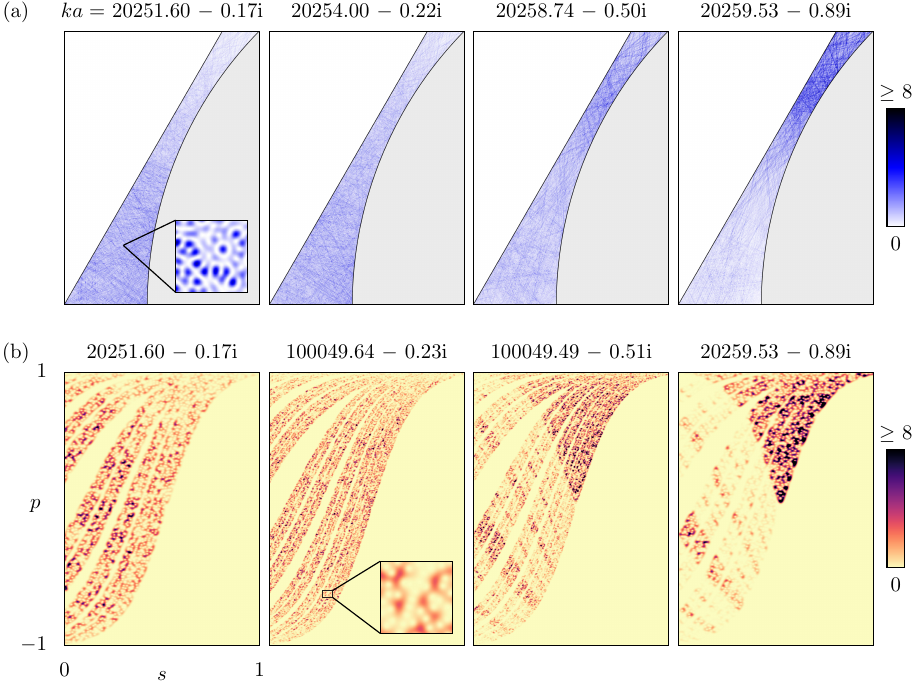}
    \caption{Resonance states for $R/a = 2.1$ at four representative values of
        $|\Imag \: ka|$, increasing from left to right.
        (a) Intensity $\psiSqVec$ in position space (fundamental domain confined
        to a region of size $0.5 a \times 0.7 a$) for $\Real \: ka \approx
        2 \cdot 10^4$.
        The inset (left) is a magnification by a factor 150 showing fluctuations
        on the scale of the wavelength.
        (b) Husimi representation $\HusVec$ on the boundary phase space $\vecx =
        (s, p)$ of a disk for $\Real \: ka \approx 2 \cdot 10^4$ (left, right)
        and even larger wavenumber $\Real \: ka \approx 10^5$ (second and third
        from left).
        The inset (second from left) shows fluctuations on the scale
        of the Planck cell.
        In all figures the average value (in (a) on fundamental domain, in (b)
        on backward-trapped set) is scaled to one  and intensities greater than
        the maximal value of the colorbar are shown with darkest color.
    }%
    \label{fig:resonance_states}
\end{figure}

\begin{figure}
    \centering
    \includegraphics{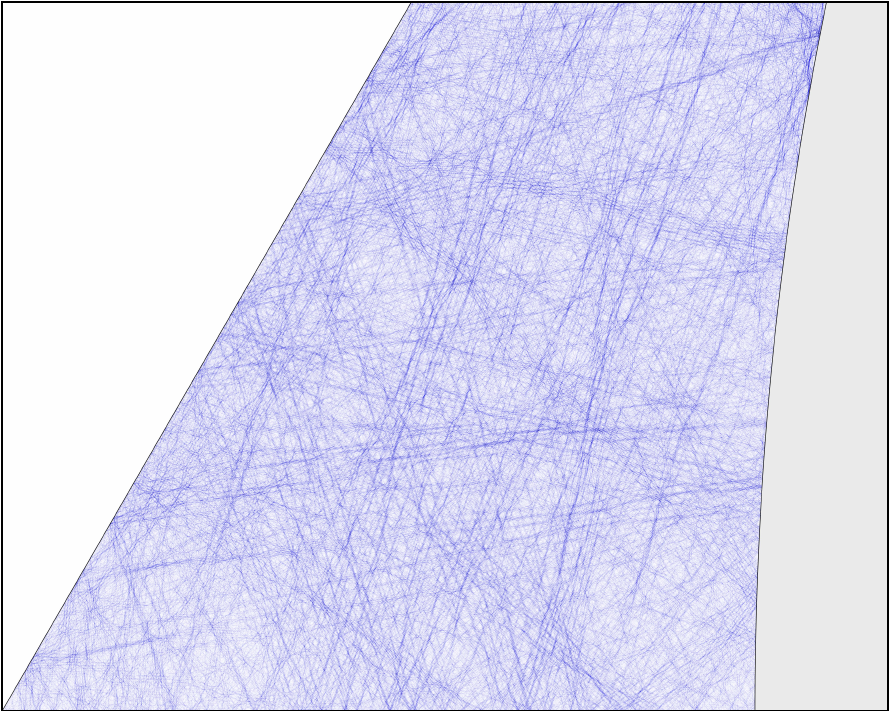}
    \caption{Ray-segment scars in a magnification of the resonance
        state from \cref{fig:resonance_states}(a, second from right) at
        $ka = 20258.74-0.50\ui$ shown for the region $0.25 a \times 0.2 a$
        (corresponding to about $800 \times 650$ wavelengths).
    }%
    \label{fig:ray_segment_scarring}
\end{figure}

In \cref{fig:resonance_states}(a) the intensity $\psiSqVec$ of four
representative resonance states is shown in position space.
Due to the large wavenumber $\Real \: ka \approx 2 \cdot 10^4$ the wave nature
of the resonance states becomes apparent only by zooming into a small region of
position space.
These resonance states show a strong dependence on $\Imag \: ka$ which
corresponds to a varying dimensionless decay rate $\gamma \tau$,
\cref{eq:decay_rate}.
For the smallest decay rate (left) the resonance state is strong in the inner
region between the three disks.
For larger decay rates (right) it becomes stronger towards the outer region of
the scattering system.

A resonance state can be represented by a Husimi function $\HusVec$ on the
boundary phase space $\vecx = (s, p)$~\cite{BaeFueSch2004, WeiBarKuhPolSch2014}.
This is shown in \cref{fig:resonance_states}(b) including two representative
examples further in the semiclassical limit at $\Real \: ka \approx 10^5$ with
similar values of $\Imag \: ka$ as in \cref{fig:resonance_states}(a).
They concentrate on the fractal backward-trapped
set~$\bwdset$~\cite{CasMasShe1999b, KeaNovPraSie2006, Dya2019}, cf.\
\cref{fig:invariant_sets}(a).
The Husimi function is smooth on a scale of area $2\pi/(\Real \: ka)$
corresponding to a Planck cell, which needs a magnification for the studied
large value of $k$.
For increasing decay rates (right) one observes more structure along the
backward-trapped set.
In particular, the intensity on the chaotic saddle is decreased.

We observe that resonance states with a similar value of $\Imag \: ka$,
i.e.\ similar dimensionless decay rate $\gamma \tau$, \cref{eq:decay_rate}, show
the same overall structure in position and Husimi representation (with
increasingly fine details for increasing $\Real \: ka$).
This observation motivates the factorization of resonance states discussed in
\cref{sec:factorization}.
For convenience we use from here on for the dimensionless decay rate
$\gamma \tau$ just the symbol $\gamma$.

In this paper we restrict the investigation
to the parameter $R/a = 2.1$ of the three-disk scattering system.
Values of $R/a$ slightly above 2 have the advantage that the backward-trapped
set $\bwdset$ has a large fractal dimension close to the maximal integer value.
As the resonance states concentrate on the backward-trapped set,
see \cref{fig:resonance_states}(b),
they can be much better analyzed for such values of $R/a$.
In particular, there are many more fluctuations in the Husimi function
than for the case of larger $R/a$, where the fractal dimension of the
backward-trapped set tends to zero.

Recently, in dielectric cavities it was observed that all resonance states
for large wavenumber $k$ show strong scars,
which were termed ray-segment scars~\cite{KetClaFriBae2022, KetClaFriBae2022:supp}.
It can be observed just as well for the three-disk scattering system
in all resonance states of \cref{fig:resonance_states}(a)
and the magnification of one of them in \cref{fig:ray_segment_scarring}.
A ray-segment scar is an enhancement of the intensity $\psiSqVec$
along a segment of a ray.
It extends for thousands of wavelengths, often beyond one or
two reflections on the boundary.
It is not related to the well-known periodic-orbit scars~\cite{Hel1984,
Kap1999, Ver2015}
of eigenstates of closed chaotic billiards,
which occur for a small fraction of eigenstates only.
In contrast, ray-segment scars are observed in every resonance state
at sufficiently large wavenumber $k$.
That ray-segment scars are not on periodic orbits can be seen best in
\cref{fig:resonance_states}(a, right) where one finds scars along rays leaving
the system.
Ray-segment scars can be conceptually explained~\cite{KetClaFriBae2022}
based on the factorization of resonance states discussed in
\cref{sec:factorization}.
Their properties will be discussed in \cref{sec:discussion}.
The observation of ray-segment scars in the
three-disk scattering system also establishes this phenomenon in systems with
full escape.
Together with their observation in dielectric cavities with partial
escape~\cite{KetClaFriBae2022}, this demonstrates the universal relevance of
ray-segment scars for quantum chaotic scattering.

\subsection{Factorization}%
\label{sec:factorization}

\begin{figure}
    \centering
    \includegraphics{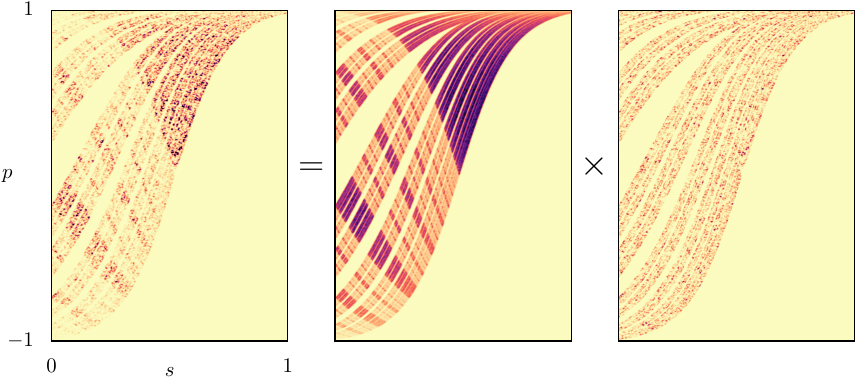}
    \caption{Factorization of an exemplary resonance state in Husimi
        representation $\HusVec$ on the boundary phase space (left) into an average
        $\HusAvgVec$ of resonance states with similar decay rate (middle) and
        fluctuations $\eta(\vecx)$ (right).
        The resonance state is taken from \cref{fig:resonance_states}(b,
        second from right) and the average is also shown in
        \cref{fig:resonance_states_avg}(b, second from right).
    }%
    \label{fig:factorization}
\end{figure}

In Refs.~\cite{ClaKunBaeKet2021, KetClaFriBae2022} the following conjecture was
stated:
\textit{Chaotic resonance states $\psi(\vecr)$ in scattering systems are a product
\begin{equation}
    \label{eq:factorization}
    |\psi(\vecr)|^2 = \rho_\gamma(\vecr) \cdot \eta(\vecr)
\end{equation}
of (i) conditionally invariant measures from classical
dynamics with a smoothed spatial density $\rho_\gamma(\vecr)$ depending on the
states’s decay rate $\gamma$ and (ii) universal exponentially distributed
fluctuations $\eta(\vecr)$ with mean one.}
The equality in \cref{eq:factorization} is understood in the sense that
the right hand side has the same statistical properties as the
resonance state on the left hand side.
This factorization is just as well conjectured to hold in the Husimi
representation,
\begin{equation}
    \label{eq:factorization_husimi}
    \Hus(\vecx) = \rho_\gamma(\vecx) \cdot \eta(\vecx),
\end{equation}
in phase space.
This was validated for quantum maps~\cite{ClaKunBaeKet2021} and dielectric
cavities~\cite{KetClaFriBae2022}.
Here, we verify it for the three-disk scattering system, an autonomous
system with full escape.

Averaging \cref{eq:factorization,eq:factorization_husimi} over resonance states
with a similar value of $\Imag \: ka$, i.e.\ similar decay rate $\gamma$, leads
to
\begin{equation}
    \label{eq:averages}
    \rho_\gamma(\vecr) = \psiAvgVec \quad \mathrm{and} \quad
    \rho_\gamma(\vecx) = \HusAvgVec,
\end{equation}
relating the classical densities to the corresponding quantum average.

Before studying the semiclassical origin of $\rho_\gamma(\vecr)$ and
$\rho_\gamma(\vecx)$ in
\cref{sec:semiclassical_structure}, we present in \cref{sec:avg_res_states}
the averages $\psiAvgVec$ and $\HusAvgVec$ and discuss their dependence on the
decay rate.
In order to study the universality of the fluctuations $\eta(\vecr)$ and
$\eta(\vecx)$ without using any knowledge about $\rho_\gamma(\vecr)$ and
$\rho_\gamma(\vecx)$,
we replace $\rho_\gamma$ in \cref{eq:factorization,eq:factorization_husimi} by
the average, \cref{eq:averages}.
This gives for the fluctuations of the resonance state $\psi$ with decay rate
$\gamma$,
\begin{equation}
    \label{eq:def_fluctuations}
    \eta(\vecr) = \frac{\psiSqVec}{\psiAvgVec} \quad \mathrm{and} \quad
    \eta(\vecx) = \frac{\HusVec}{\HusAvgVec} \quad
    (\vecx \in \bwdset),
\end{equation}
where $\vecx$ has to be chosen on the backward-trapped set $\bwdset$.
These fluctuations will be studied in \cref{sec:fluctuations}.

The factorization of an exemplary resonance state in the Husimi representation,
$\HusVec = \HusAvgVec \cdot \eta(\vecx)$, is visualized in
\cref{fig:factorization}.
The average shows a clear multifractal structure
and the fluctuations show a uniform distribution on the backward-trapped set.

\subsection{Average resonance states}%
\label{sec:avg_res_states}

\begin{figure}
    \centering
    \includegraphics{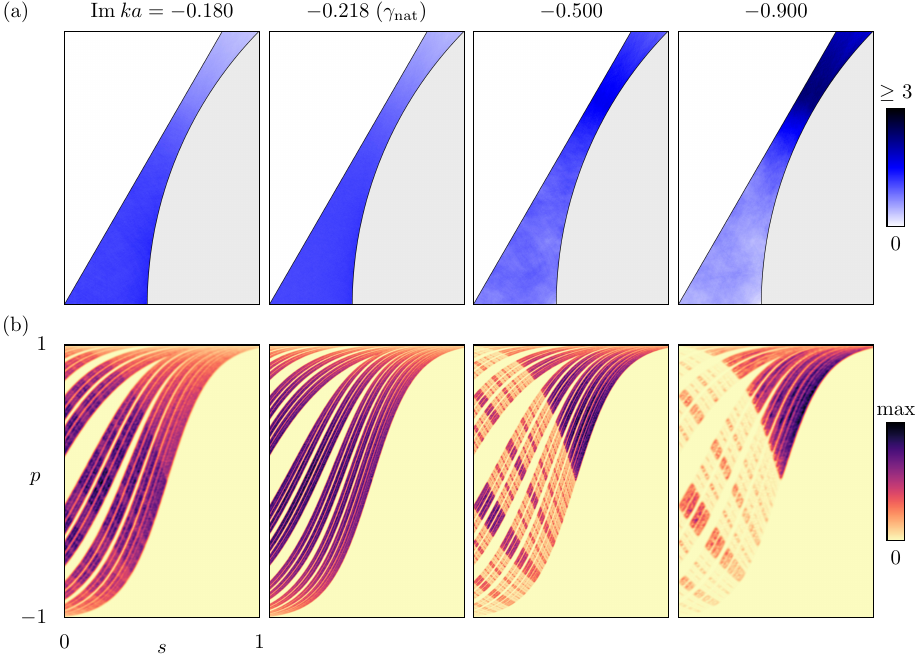}
    \caption{Averages $\psiAvgVec$ and $\HusAvgVec$ over 500 resonance states
        with imaginary part closest to four representative values of $\Imag \:
        ka$ in (a) position space where $\Real \: ka \in [20000, 20500]$
        and (b) Husimi representation where $\Real \: ka \in [100000,
        100100]$ (second and third from left) and $\Real \: ka \in [20000,
        20500]$ (left, right), otherwise corresponding to \cref{fig:resonance_states}.
        The contributing resonance poles are highlighted in \cref{fig:spectrum}.
        In (b) the maxima are approximately given by 2.1, 1.8, 2.9, and 5.6 (from
        left to right).
    }%
    \label{fig:resonance_states_avg}
\end{figure}

The averages in position representation $\psiAvgVec$ and on the boundary phase space
$\HusAvgVec$ at four representative value of $\Imag \: ka$ are shown in
\cref{fig:resonance_states_avg}.
The averaging is in each case done over 500 resonance states with poles closest
to the chosen $\Imag \: ka$.
These resonance poles are also highlighted in \cref{fig:spectrum}.

For the average of resonance states their normalization relative to each other
is important.
While for closed billiards~\cite{Rel1940, BerWil1984, Boa1994} and dielectric
cavities~\cite{KetClaFriBae2022} this is well-defined by integrating the
absolute square of the wave function over the area of the system, there is no
obvious area to integrate over for the three-disk scattering system.
Therefore, we choose to normalize the resonance states such that
$\int_0^{2 \pi} \ud s \; |a^2 \phi_n(s)|^2 = |k_n a|^2$
where $\phi_n(s)$ is the normal derivative of the wave function $\psi_n(\vecr)$
on the outer side of the disk's boundary.
This is motivated by the normalization of a wave function inside a closed
circular boundary~\cite{Rel1940, BerWil1984, Boa1994}
(up to an irrelevant factor of 2).
In particular, it is numerically convenient, see \cref{sec:appendix_position},
in contrast to integrating $|\psi_n(\vecr)|^2$ over some region in position
space.

In \cref{fig:resonance_states_avg}(b) the averaged resonance states in
Husimi representation show a
strong dependence on $\Imag \: ka$, which is even more pronounced than for
individual resonance states, see \cref{fig:resonance_states}(b).
As expected for $\Imag \: ka$ corresponding to the natural decay rate $\gnat$
(\cref{fig:resonance_states_avg}(b), second from left) a
uniform distribution on the backward-trapped set is
observed~\cite{LeeRimRyuKwoChoKim2004, ShiHar2007, WieHen2008,
ShiHenWieSasHar2009, ShiHarFukHenSasNar2010, HarShi2015, KulWie2016,
BitKimZenWanCao2020, KetClaFriBae2022} corresponding to the natural measure
$\munat$, see \cref{fig:invariant_sets}(d).
By smoothing $\munat$ on the scale of a Planck cell, see
\cref{fig:husimi_ray_comparison} (second from left), perfect agreement is
achieved.
With increasing $|\Imag \: ka|$ (\cref{fig:resonance_states_avg}(b), to the
right) the probability density on the backward-trapped set becomes fractal along
the backward-trapped set, in particular it decreases on the chaotic saddle.
For resonance states with smallest $|\Imag \: ka|$
(\cref{fig:resonance_states_avg}(b), left) one observes a small increase
on the chaotic saddle, which is noticeable by averaging only.
The multifractal structure is qualitatively well described by the partitioning
of the backward-trapped set based on the stable and unstable manifold of the
periodic orbit at the exit, see \cref{fig:invariant_sets}(f).

In position representation, shown in \cref{fig:resonance_states_avg}(a), the
averaged resonance states, just like individual resonance states, become
stronger towards the outer region of the scattering system for increasing
$|\Imag \: ka|$.
Beyond that one finds very weak structure especially in comparison to dielectric
cavities in position representation~\cite{KetClaFriBae2022}.
In \cref{sec:semiclassical_structure} we will relate the averages $\psiAvgVec$
and $\HusAvgVec$ according to \cref{eq:averages} to classical densities
$\rho_\gamma(\vecr)$ and $\rho_\gamma(\vecx)$.

\subsection{Universal fluctuations}%
\label{sec:fluctuations}

\begin{figure}[t]
    \centering
    \includegraphics{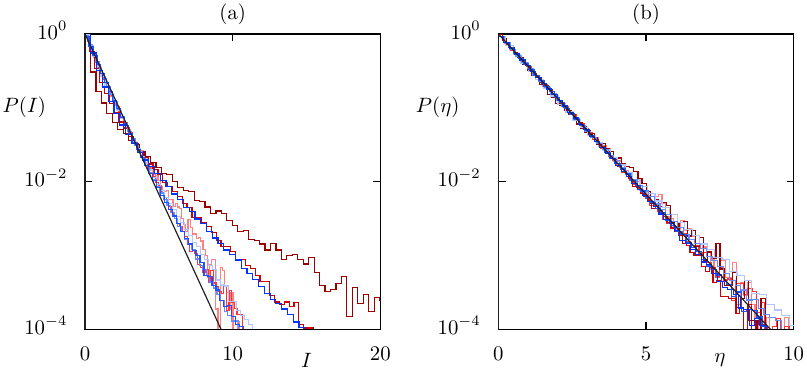}
    \caption{Distribution of (a) the intensities $I = |\psi(\vecr)|^2$ (blue)
        and $I = \Hus(\vecx), \: x \in \bwdset$ (red) and (b) the fluctuations
        $\eta(\vecr)$ (blue) and $\eta(\vecx)$ (red) for the resonance states of
        \cref{fig:resonance_states}, colored by increasing $|\Imag \: ka|$
        from light to dark.
        For comparison an exponential distribution with mean one is shown
        (black).
    }%
    \label{fig:universal_fluctuations}
\end{figure}

In closed systems the distribution of the intensities $\psiSqVec$ and $\HusVec$
of chaotic eigenstates follow the universal distribution of the random wave
model~\cite{Ber1977b, McDKau1988, AurSte1991, LiRob1994, Pro1997b}.
For the complex (real) random wave model this is a complex (real) Gaussian
distribution of the eigenstate amplitudes leading to an exponential
(Porter-Thomas) distribution of the intensities.

In \cref{fig:universal_fluctuations}(a) the distribution of the intensities
$|\psi(\vecr)|^2$ and the Husimi representation $\HusVec$ for $\vecx
\in \bwdset$ are shown for the resonance states from
\cref{fig:resonance_states}.
One finds different distributions for different $\Imag \: ka$ and for the two
representations, in particular no exponential distribution.

In \cref{fig:universal_fluctuations}(b) the distributions of the
fluctuations $\eta(\vecr)$ and $\eta(\vecx)$, defined in
\cref{eq:def_fluctuations}, are shown.
They follow a universal exponential distribution with mean one over more than
three orders of magnitude.
This shows that the fluctuations $\eta$ follow a complex random wave model,
demonstrating the importance of the factorization for understanding the
structure of resonance states.
This is in line with previous findings for quantum maps~\cite{ClaKunBaeKet2021}
and dielectric cavities~\cite{KetClaFriBae2022}.

\section{Semiclassical structure of average resonance states}%
\label{sec:semiclassical_structure}

After demonstrating the factorization in \cref{sec:factorization_states},
we now want to relate the averages $\psiAvgVec$ and $\HusAvgVec$ according to
\cref{eq:averages} to classical densities $\rho_\gamma(\vecr)$ and
$\rho_\gamma(\vecx)$.
These densities are obtained from smoothed conditionally-invariant
measures~\cite{DemYou2006} with decay rate $\gamma$, in analogy to limit
measures of sequences of individual resonance states~\cite{NonRub2007}.
In particular, for the natural decay rate $\gnat$ the natural measure
$\munat$ describes resonance states of this decay
rate~\cite{LeeRimRyuKwoChoKim2004, ShiHar2007, WieHen2008,
ShiHenWieSasHar2009, ShiHarFukHenSasNar2010, HarShi2015, KulWie2016,
BitKimZenWanCao2020, KetClaFriBae2022}.
Indeed, smoothing $\munat$ from \cref{fig:invariant_sets}(d) on the scale of a
Planck cell in phase space yields perfect agreement, see
\cref{fig:husimi_ray_comparison} (second from left).
This also holds in position space, see \cref{fig:position_ray_comparison}
(second from left).

For other decay rates we adopt a conditionally-invariant measure from
quantum maps in \cref{sec:measure_temporal_distance}, which very well
describes the averaged resonance states.
We find in \cref{sec:jensen_shannon}, however, that this
is not the correct semiclassical limit measure.
Recent improvements are discussed in \cref{sec:discussion}.

\subsection{Measure based on temporal distance}%
\label{sec:measure_temporal_distance}

In a system with full escape it was conjectured in Ref.~\cite{ClaKoeBaeKet2018}
that the semiclassical measure is uniformly distributed on sets with the same
temporal distance $\tdist$ to an $\epsilon$-surrounding of the chaotic saddle
$\chsaddle$.
This yields a conditionally-invariant measure with decay rate~$\gamma$,
\begin{equation}
    \label{eq:scl_measure}
    \mu^\epsilon_{\gamma}(A) \propto \int_A
    \ue^{t_\epsilon(\vecx) (\gamma - \gnat)} \: \ud
    \munat(\vecx)
\end{equation}
for any set $A$ in phase space.
This measure modifies the natural measure $\munat$ depending on the temporal
distance $\tdist(\vecx)$ shown in \cref{fig:invariant_sets}(d, e).
The modification becomes stronger with increasing deviation of the desired decay
rate $\gamma$ from $\gnat$.
For the reasoning behind \cref{eq:scl_measure}, see
Ref.~\cite{ClaKoeBaeKet2018}.

\begin{figure}
    \centering
    \includegraphics{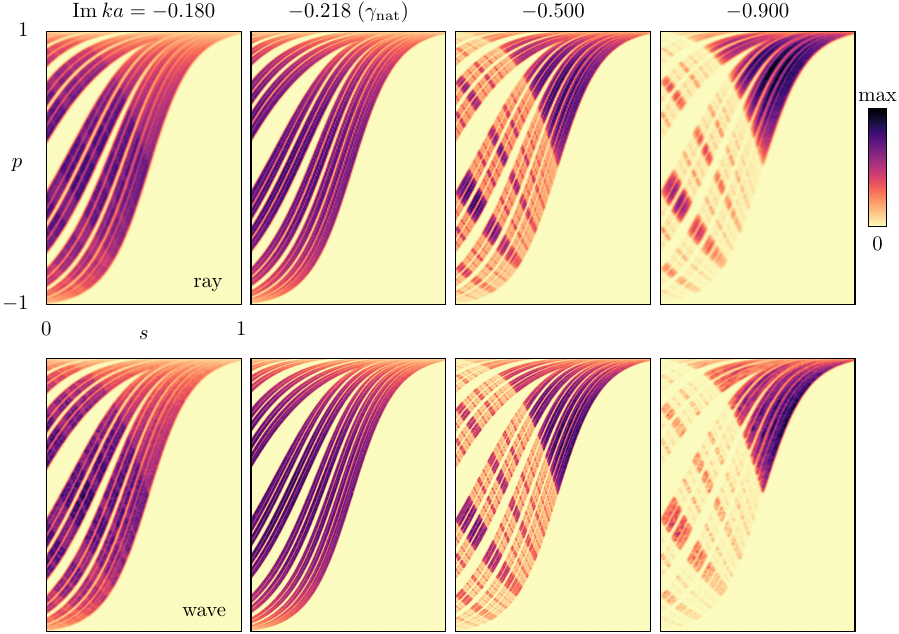}
    \caption{Smoothed measure $\mu^\epsilon_{\gamma}$ in phase space for
        $\epsilon = 10^{-4}$ (ray, top) compared to averaged resonance states
        $\HusAvgVec$ (wave, bottom), repeated from
        \cref{fig:resonance_states_avg}(b) for the convenience of the reader.
    }%
    \label{fig:husimi_ray_comparison}
\end{figure}

The measure $\mu^\epsilon_{\gamma}$ smoothed on the scale of the corresponding
Planck cell is displayed in phase space in \cref{fig:husimi_ray_comparison}
and compared to the averaged Husimi representation $\HusAvg$.
On a qualitative level, we observe very good agreement for the three
additional decay rates $\gamma \neq \gnat$.
In particular, this holds for the partitioning by the stable and unstable
manifold of the periodic orbit at the exit, see
\cref{fig:invariant_sets}(f).
However, we observe that the value of the measure shows some deviations which
increase with the distance from $\gnat$.
This will be further quantified using the Jensen-Shannon divergence, see
\cref{sec:jensen_shannon}.

\begin{figure}
    \centering
    \includegraphics{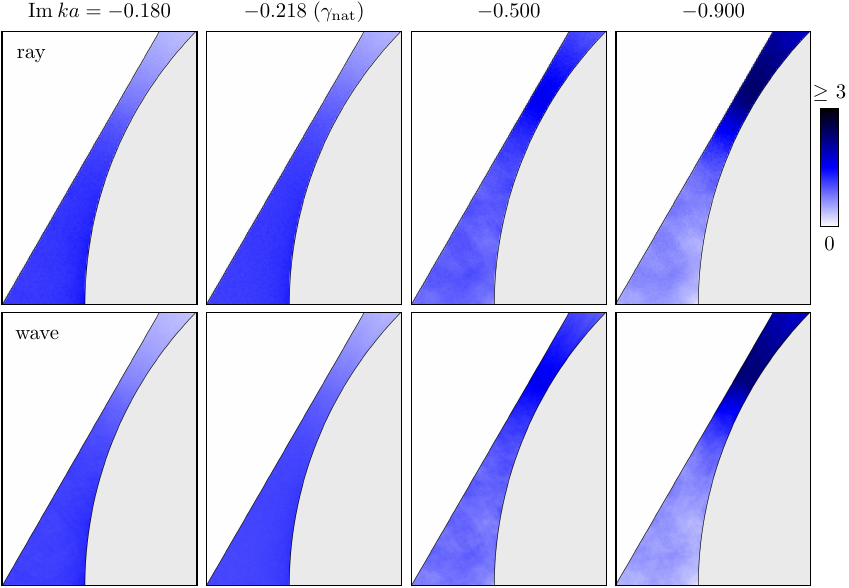}
    \caption{Smoothed measure $\mu^\epsilon_{\gamma}$ in position space for
        $\epsilon = 10^{-4}$ (ray, top) compared to averaged resonance states
        $\psiAvgVec$ (wave, bottom), repeated from
        \cref{fig:resonance_states_avg}(a) for the convenience of the reader.
    }%
    \label{fig:position_ray_comparison}
\end{figure}

In position space one can define a measure $\mu^\epsilon_{\gamma}(\vecr)$ in
analogy to \cref{eq:scl_measure}.
It is shown in \cref{fig:position_ray_comparison}, integrated over the size
of each pixel, and compared to the averaged intensities $\psiAvgVec$.
We find very good agreement, with the grain of salt that there is not much
structure in position space to make a detailed comparison.

Note that in Ref.~\cite{ClaKoeBaeKet2018} for the $\epsilon$-surrounding the
specific value $\epsilon = \sqrt{\hbar/2}$ was used, leading to the
\textit{$h$-resolved chaotic saddle}.
In our case of $\Real \: ka \approx 10^5$ this would correspond to $\epsilon =
\sqrt{1/(2 \: \Real \: ka)} \approx 2 \cdot 10^{-3}$, which however does not
lead to a good agreement in a figure (not shown) equivalent to
\cref{fig:husimi_ray_comparison}, where we choose heuristically $\epsilon =
10^{-4}$.
We explain this in the following way:
In order to have a certain resolution of the temporal distance $\tdist$ along
the backward-trapped set (i.e.\ the unstable direction) one needs to define the
$\epsilon$-surrounding of the chaotic saddle on a much finer scale in this
direction.
As the temporal distance has a sufficiently fine resolution, see
\cref{fig:invariant_sets}(e), the chosen value of $\epsilon = 10^{-4}$ seems
sufficiently small.
In fact for $\epsilon = 10^{-5}$ we find no changes.

\subsection{Jensen-Shannon divergence}%
\label{sec:jensen_shannon}

\begin{figure}
    \centering
    \includegraphics{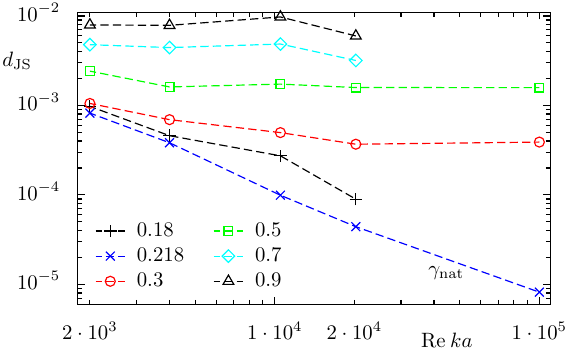}
    \caption{Jensen-Shannon divergence $d_\mathrm{JS}$ between the measure
        $\mu_\gamma^\epsilon$ and the averaged Husimi representations $\HusAvg$
        over 500 resonance states computed for $\Real \: ka \in [1000, 3000], \,
        [3000, 5000], \, [10000, 11000], \, [20000, 20500], \, [100000,
        100100]$, and $|\Imag \: ka|$ near \{0.218 ($\gnat$), 0.18, 0.3, 0.5,
        0.7, 0.9\} (from bottom to top).
        At $\Real \: ka \approx 10^5$ only some data points are shown, as there
        are no poles near $|\Imag \: ka| = 0.18$ and no resonance states have
        been computed for $|\Imag \: ka| \geq 0.7$.
    }%
    \label{fig:jensen_shannon}
\end{figure}

In \cref{fig:husimi_ray_comparison} we find for the averaged Husimi
representation perfect agreement for $\gnat$ but observe deviations for $\gamma
\neq \gnat$ using the measure $\mu^\epsilon_\gamma$ based on the temporal
distance.
We now quantify this using the Jensen-Shannon divergence
$d_\mathrm{JS}$~\cite{Lin1991}, which compares two probability distributions.
This was used to compare a semiclassical measure to resonance states in systems
with partial escape in Ref.~\cite{ClaAltBaeKet2019}.
For computing the Jensen-Shannon divergence one has to choose a partitioning of
phase space.
For studying the semiclassical limit we choose a fixed scale independent of $k$
given by squares with side length $\sqrt{h_\mathrm{max}} = \sqrt{2\pi/(\Real \:
k_{\mathrm{min}}a)}$ and $k_{\mathrm{min}} a = 1000$.

The Jensen-Shannon divergence is shown in \cref{fig:jensen_shannon} for
six decay rates and $\Real \: ka$ from $10^3$ to $10^5$.
For $\gnat$, corresponding to $|\Imag \: ka| = 0.218$, we find convergence in
the semiclassical limit confirming the qualitative impression of
\cref{fig:husimi_ray_comparison} and previous
findings~\cite{LeeRimRyuKwoChoKim2004, ShiHar2007, WieHen2008,
ShiHenWieSasHar2009, ShiHarFukHenSasNar2010, HarShi2015, KulWie2016,
BitKimZenWanCao2020, KetClaFriBae2022}.
For $\gamma < \gnat$, corresponding in \cref{fig:jensen_shannon} to
$|\Imag \: ka| = 0.18$, one finds small values of $d_\mathrm{JS}$, however the
convergence cannot be studied, as there are no resonance poles in this region in
the semiclassical limit, see \cref{fig:spectrum}.
For $\gamma > \gnat$ we find no convergence in the semiclassical limit which
suggests that the measure $\mu^\epsilon_\gamma$ based on the temporal distance
is not the correct semiclassical limit measure.
Recent improvements are discussed in \cref{sec:discussion}.

\section{Discussion and outlook}%
\label{sec:discussion}

We demonstrate that the factorization conjecture holds for resonance states in
the three-disk scattering system.
One factor is described by universal exponentially distributed intensity
fluctuations.
For the other factor, also describing averaged resonance states, we apply a
construction from maps with full escape to the three-disk scattering system.
We observe very good, but not perfect, agreement.
The factorization conjecture complements the periodic-orbit approach
to resonance states of the three-disk scattering system.
It allows for obtaining new insights into their structure, in particular
about their dependence on the decay rate as well as about similarities and
differences of neighboring resonance states.
By computing resonance states further in the semiclassical limit than before, we
are able to validate the fractal Weyl law over a very large range of wavenumbers.

The recently observed ray-segment scars in dielectric
cavities~\cite{KetClaFriBae2022} are also found in the three-disk scattering
system.
This demonstrates the universal relevance of ray-segment scars for quantum
chaotic scattering.
They have been explained~\cite{KetClaFriBae2022} based on the factorization conjecture:
Whenever the multifractal classical density shows strong intensity enhancements
in phase space, then the additional universal fluctuations give rise to some
phase-space points with extremely high intensities.
In every resonance state this leads to scars along
segments of rays, which are unrelated to periodic orbits.
The most likely directions are determined by the high intensities of the
multifractal classical density.
The specific direction of the ray segment varies from state to state, as the
phase-space points with extreme intensities vary due to the universal
fluctuations.
Going to the semiclassical limit the multifractal classical density is
resolved on finer scales, leading to higher intensities and thus to stronger
scars.
In the future it is desirable to analyze ray-segment scars quantitatively, in
particular their length, width and intensity distributions.

Another future aim is to derive the semiclassical measure which perfectly
describes the resonance states in systems with full or partial escape.
In particular, it would be desirable to find a common approach, as currently one
either needs the natural measure and the temporal distance (full
escape~\cite{ClaKoeBaeKet2018}) or the natural and the inverse natural measure
(partial escape~\cite{ClaAltBaeKet2019, KetClaFriBae2022}).
In fact, work in progress shows that such a common approach to derive a
semiclassical measure exists, using ideas from systems with local
randomization~\cite{ClaKet2022},
which describes the resonance states even better~\cite{KetLorSch2023:p}.
Another approach for the three-disk scattering system is
to use the semiclassical theory based on dynamical zeta functions consisting of
periodic orbits.
This may allow for a derivation of the semiclassical measure.
More generally, the periodic-orbit approach might be able to give support to the
factorization conjecture.

\appendix
\ack

We thank A.\ Bäcker, S.\ Barkhofen, F.\ Bönisch, J.\ Fleck, F.\ Lorenz,
J.\ Stöber and T.\ Weich for valuables discussions.
The authors are grateful to the Center for Information Services and High
Performance Computing (ZIH) at TU Dresden for providing its facilities for high
throughput calculations.
Funded by the Deutsche Forschungsgemeinschaft (DFG, German Research
Foundation) -- 262765445.

\section{Numerical method for computing resonances}%
\label{sec:appendix_numerical_method}

In this appendix we describe the numerical methods used for computing
resonance poles and resonance states of the three-disk scattering system.
Python code is provided as supplementary material.
In \cref{sec:appendix_matrix_vector} we describe an efficient method for
applying the matrix $M(k)$ to a vector which is used in
\cref{sec:appendix_compute_poles} to determine its zeros and the corresponding
states.
Those states are used to compute the position representation of resonance states
in \cref{sec:appendix_position} and the Husimi representation on the
boundary phase space of a disk in \cref{sec:appendix_husimi}.
We omit derivations for brevity, but rather concentrate on the equations
necessary for an implementation of the methods.

\subsection{Efficient matrix-vector multiplication}%
\label{sec:appendix_matrix_vector}

The resonance poles $k_n$ are the zeros of a matrix $M(k)$, i.e.\
$\det(M(k_n))~=~0$, which is in the $\mathrm{A_2}$-representation given
by~\cite{GasRic1989c},
\begin{eqnarray}
    \label{eq:appendix_m_matrix}
    \fl M_{m m'}(k) = \frac{\pi a}{2 \ui} \biggl\{\delta_{m m'} + 2
        \frac{J_m(ka)}{H^{(1)}_{m'}(ka)} \biggl[& H^{(1)}_{m-m'}(kR)
            \cos\left(\frac{\pi}{6}(5m-m')\right) \nonumber \\
            &- (-1)^{m'} H^{(1)}_{m+m'}(kR) \cos\left(\frac{\pi}{6} (5m+m')\right)
        \biggr]\biggr\},
\end{eqnarray}
where $m, m' \in \{1, 2, \ldots\, m_\mathrm{max}\}$ and $J_m$ $(H^{(1)}_m)$ is a
Bessel (Hankel) function of the first kind and $m$th order.
The matrix is truncated at size $m_\mathrm{max}$ of order $ka$, as larger matrix
elements are exponentially small~\cite{GasRic1989c}.

Below we can employ efficient methods for eigenproblems and linear systems, as
we find that the matrix-vector multiplication,
\begin{equation}
    \psi'_m = \sum_{m'=1}^{m_\mathrm{max}} M_{m m'}(k) \: \psi_{m'},
\end{equation}
can be very fast evaluated using fast Fourier transformations,
\begin{equation}
    \label{eq:matrix_vector_multiplication}
    \psi'_m = \frac{\pi a}{2 \ui} \Biggl\{\psi_m
        + (-1)^m  \widetilde{J}_m(ka) \:
        \mathcal{F}^{-1}\Biggl(
            f_{i + \frac{N}{12}} \:
            g_{i + \frac{N}{6}}
        +
            f_{i - \frac{N}{12}} \:
            g_{i - \frac{N}{6}}
        \Biggr)_m
    \Biggr\},
\end{equation}
where
\begin{equation}
    \label{eq:fourier_transforms}
    f_i = \mathcal{F}\left(\widetilde{H}^{(1)}_j(kR)\right)_i
    \quad \mathrm{and} \quad
    g_i =
    \mathcal{F}\left(\frac{\widetilde{\psi}_j}{\bar{H}^{(1)}_j(ka)}\right)_i
\end{equation}
are Fourier transforms $\mathcal{F}$ of vectors with indices $j \in
\{-m_\mathrm{max}, \ldots, m_\mathrm{max}\}$ zero-padded to a length $N$, which
is a multiple of 12, and giving a vector of length $N$ indexed by $i$.
The index shifts in \cref{eq:matrix_vector_multiplication} are done cyclicly and
$\mathcal{F}^{-1}$ is the inverse Fourier transform.
Furthermore, we define the antisymmetric vector $\widetilde{\psi}_j$ by
$\widetilde{\psi}_m = \psi_m$, $\widetilde{\psi}_0 = 0$, and $\widetilde{\psi}_{-m} =
-\psi_m$.
Although the usual Bessel and Hankel functions could be used in
\cref{eq:matrix_vector_multiplication,eq:fourier_transforms},
we define symmetric functions
\begin{eqnarray}
    \widetilde{J}_m &= J_m \ue^{-\ui \frac{\pi}{2} m}, \\
    \widetilde{H}^{(1)}_m &= H^{(1)}_m \ue^{\ui \frac{\pi}{2} m}, \\
    \bar{H}^{(1)}_m &= H^{(1)}_m \ue^{-\ui \frac{\pi}{2} m},
\end{eqnarray}
such that all Fourier transforms can be replaced by (discrete type I) cosine or
sine transforms of half the length.
The general idea in deriving \cref{eq:matrix_vector_multiplication} is to
interpret Eq.~(\ref{eq:appendix_m_matrix}) as a convolution and to replace the
cosine by exponential terms which are shifts of the Fourier transform.
An analogous procedure can be used to multiply the first and second derivative
of the matrix $M(k)$ with a vector $\psi$, which will be needed below.

\subsection{Computation of resonance poles and states}%
\label{sec:appendix_compute_poles}

The resonance poles $k_n$ are the zeros of the matrix $M(k)$, i.e.\
$\det(M(k_n))~=~0$~\cite{GasRic1989c}.
They will be first approximated using a method introduced in
Ref.~\cite{VebProRob2007} and afterwards iteratively converged to the desired
accuracy, in both steps using the matrix-vector multiplication from
\cref{sec:appendix_matrix_vector}.
The corresponding right and left eigenstates of $M$ with eigenvalue zero fulfill
\begin{equation}
    \label{eq:M_matrix_multiplication}
    M(k_n) | R_n \rangle = 0
    \quad \mathrm{and} \quad
    \langle L_n | M(k_n) = 0.
\end{equation}
They are necessary for the left and right resonance states of the three-disk
scattering system, see \cref{sec:appendix_position}.

Approximate poles $k_n$ and eigenstates $|R_n \rangle$, $\langle L_n |$ are
determined in the neighborhood of some $\kappa \in \mathbb{C}$ by (for
derivation see Refs.~\cite{VebProRob2007, PeiDieHua2019})
\begin{equation}
	k_n = \kappa + k_n^{(0)} + k_n^{(1)},
\end{equation}
where the $k_n^{(0)}$ are determined from the generalized eigenvalue problem
\begin{eqnarray}
    &M(\kappa) \: |R_n \rangle  = -k^{(0)}_n &M'(\kappa) \: |R_n \rangle,%
    \label{eq:generalized_eigenvalue_problem1} \\
    \langle L_n | \: &M(\kappa) \phantom{\: |R_n \rangle}
    = - k_n^{(0)} \; \langle L_n | \: &M'(\kappa),%
    \label{eq:generalized_eigenvalue_problem2}
\end{eqnarray}
with $M'(\kappa)$ being the derivative of $M(k)$ at $k = \kappa$.
The correction term $k_n^{(1)}$ is given by
\begin{equation}
	\label{eq:first_order_correction}
	k_n^{(1)}
	=
	- \frac{(k_n^{(0)})^2}{2}
	\frac{\langle L_n | M''(\kappa) | R_n \rangle}
	{\langle L_n | M' (\kappa) | R_n \rangle} \,.
\end{equation}
We apply this method, used for real $k_n$ in Refs.~\cite{VebProRob2007,
PeiDieHua2019}, to complex $k_n$, as previously done for dielectric
cavities~\cite{KetClaFriBae2022}.
We use a neighborhood of $\kappa$ of width $\Real(\Delta ka) = 0.5$ and
$\Imag(\Delta ka) = 1$ giving about 100 poles for $ka = 10^5$ and $R/a = 2.1$.

To obtain those poles and corresponding states we approximately solve the
generalized eigenvalue problem,
Eqs.~(\ref{eq:generalized_eigenvalue_problem1}) and
(\ref{eq:generalized_eigenvalue_problem2}),
using the matrix-vector multiplication from \cref{sec:appendix_matrix_vector} in
the following way:
We determine the 1\% eigenvalues of smallest magnitude of $M(\kappa)$,
\begin{equation}
    M(\kappa) | r_i \rangle = \xi_i | r_i \rangle, \quad
    \langle \ell_i | M(\kappa) = \xi_i \langle \ell_i |,
\end{equation}
using the Arnoldi method for right and left eigenstates individually.
Those are used to project the generalized eigenvalue problem,
Eqs.~(\ref{eq:generalized_eigenvalue_problem1}) and
(\ref{eq:generalized_eigenvalue_problem2}),
to a much smaller subspace.
In this subspace we define the (small) matrix
\begin{equation}
    A_{ji} = \frac{1}{\xi_j \langle \ell_j | r_j \rangle}
        \langle \ell_j | M'(\kappa) | r_i \rangle,
\end{equation}
and solve its eigenproblem (by standard diagonalization)
\begin{equation}
    \sum_i A_{ji} \: a_{in} = \lambda_n \: a_{jn} \,, \quad
    \sum_i {(A^\dagger)}_{ji} \: b_{in} = \lambda^\dagger_n \: b_{jn},
\end{equation}
where $^\dagger$ denotes Hermitian conjugation.
This leads to the solutions
\begin{equation}
    k_n^{(0)} = -\frac{1}{\lambda_n}, \quad
    |R_n \rangle = \sum_i a_{in} | r_i \rangle, \quad
    |L_n \rangle = \sum_i
        \frac{1}{(\xi_i \langle \ell_i | r_i \rangle)^\dagger} \:
        b_{in} | \ell_i \rangle,
\end{equation}
of the original generalized eigenvalue problem,
Eqs.~(\ref{eq:generalized_eigenvalue_problem1}) and
(\ref{eq:generalized_eigenvalue_problem2}), and $k_n^{(1)}$ can be evaluated by
\cref{eq:first_order_correction}.

We improve each pole $k_n$ and its right and left eigenstates,
$| R_n \rangle$ and $| L_n \rangle$, iteratively.
The following set of ($m_\mathrm{max} + 1$) linear equations has to be solved
for each pole $k$ (dropping the index $n$ for clarity),
\begin{equation}
    \left(\begin{array}{@{}cc@{}}
		M(k)         & M'(k) \, | R \rangle  \\
		\langle R |  & 0
    \end{array}\right)
    \left(\begin{array}{@{}c@{}}
		| \delta R \rangle   \\
		\delta k
    \end{array}\right)
	=
    \left(\begin{array}{@{}c@{}}
		- M(k) \, | R \rangle   \\
		0
    \end{array}\right)
	\; ,
\end{equation}
using a numerical method which makes use of the matrix-vector multiplication
from \cref{sec:appendix_matrix_vector}.
Here, the first set of equations is a Taylor expansion of $M(k)$ and
$| R \rangle$ in \cref{eq:M_matrix_multiplication} to lowest order and the
second equation ensures orthogonality of the correction to $| R \rangle$.
This gives an improved pole and right eigenstate
\begin{equation}
    k' = k + \delta k, \quad
    | R' \rangle = | R \rangle + | \delta R \rangle.
\end{equation}
Similarly, the set of linear equations
\begin{equation}
    \left(\begin{array}{@{}cc@{}}
		M^\dagger(k) & M'^\dagger(k) \, | L \rangle  \\
		\langle L |  & 0
    \end{array}\right)
    \left(\begin{array}{@{}c@{}}
		| \delta L \rangle   \\
		\delta k^\dagger
    \end{array}\right)
	=
    \left(\begin{array}{@{}c@{}}
		- M^\dagger(k) \, | L \rangle   \\
		0
    \end{array}\right)
	\; ,
\end{equation}
improves the pole and the left eigenstate
\begin{equation}
    k' = k + \delta k, \quad
    | L' \rangle = | L \rangle + | \delta L \rangle.
\end{equation}
These steps are repeated until the corrections are on the order of the desired
accuracy.
Previously, this method of convergence has been used by one of the authors for
dielectric cavities~\cite{KetClaFriBae2022}.

\subsection{Position representation}%
\label{sec:appendix_position}

The position representation of the (right) resonance state is given
by~\cite{WeiBarKuhPolSch2014},
\begin{equation}
    \label{eq:appendix_wavefunc}
    \psi_n(\vecr) = -\frac{\pi a}{2\ui}
    \sum_{j=1}^3 \sum_{m=-\infty}^{\infty}
    A_{jm;n} H_m^{(1)}(k_n r_j) J_m(k_n a) \ue^{\ui m \theta_j}
\end{equation}
where $(r_j, \theta_j)$ denotes the point $\vecr$ in local polar coordinates of
disk $j$ (relative to the symmetry line through disk $j$, see
Ref.~\cite[Fig.~2]{WeiBarKuhPolSch2014}).
It is expressed in terms of the Fourier coefficients $A_{jm;n}$ of the
(dimensionless) normal derivative of the $n$th wave function
$\psi_n(\vecr)$ on the boundary of disk $j$,
$\boldsymbol{n}(\theta_j) \cdot \nabla \psi_n(r_j \! = \! a, \theta_j)
\, a^2 = \sum_m A_{jm;n} \exp(\ui m \theta_j)$.
In $\mathrm{A}_2$-representation one has for all disks
$A_{1m;n} = A_{2m;n} = A_{3m;n} = A_{m;n}$
and antisymmetry in $m$,
$A_{j,-m;n} = -A_{jm;n}$ (in particular $A_{j0;n} = 0$)~\cite{GasRic1989c}.
The $A_{m;n}$ for $m \in \{1, \ldots, m_\mathrm{max}\}$ are given by
$A_{m;n} = \langle m | L_n \rangle$,
i.e.\ the $m$th component of the left eigenstate of the matrix $M_{m m'}$,
Eq.~(\ref{eq:appendix_m_matrix}).

The sum over $m \in \{-m_\mathrm{max}, \ldots, m_\mathrm{max}\}$ in
\cref{eq:appendix_wavefunc} can be computed efficiently on an equidistant grid
of $\theta$ values by one inverse fast Fourier transform (or a type 1 discrete
sine transform of half the length by using the symmetries).
This is repeated on an equidistant grid in radius $r$, giving the summand on a
polar grid $(r, \theta)$ for one disk.
For each point $\boldsymbol{r}$ with local polar coordinates $(r_j, \theta_j)$
with respect to each disk $j$ the contribution to $\psi_n$ is evaluated by 2D
spline interpolation and summed over $j$.

The normalization, $\int_0^{2 \pi} \ud s \; |a^2 \phi_n(s)|^2 = |k_n a|^2$,
given in the main text can be expressed by the Fourier coefficients $A_{m;n}$ of
$\phi_n$ yielding
\begin{equation}
    \sum_m |A_{m;n}|^2 = \frac{1}{2\pi} |k_n a|^2.
\end{equation}

\subsection{Husimi representation}%
\label{sec:appendix_husimi}

The Husimi representation of the (right) resonance state on the boundary phase
space of disk $j$ in Birkhoff coordinates $(s, p)$ is given
by~\cite{WeiBarKuhPolSch2014},
\begin{equation}
    \label{eq:appendix_husimi}
    \mathcal{H}_n(s, p) \propto \Bigg|\sum_{m=-\infty}^{\infty} A_{jm;n} \ue^{\ui m s}
    \exp\Biggl(-\frac{{(p \, \Real(k_n a) - m)}^2}{2 \Real(k_n a)}\Biggr)\Bigg|^2,
\end{equation}
where in the $\mathrm{A}_2$-representation $A_{jm;n} = A_{m;n}$.
The sum over $m \in \{-m_\mathrm{max}, \ldots, m_\mathrm{max}\}$ in
\cref{eq:appendix_husimi} can be computed efficiently on an equidistant grid
of $s \in [-\pi, \pi]$ by one inverse fast Fourier transform for each $p \geq
0$.
For $p < 0$ holds $\mathcal{H}_n(s, p) = \mathcal{H}_n(-s, -p)$.

\end{document}